\documentclass[aps,pra,twocolumn,groupedaddress,showpacs,showkeys]{revtex4-1}
\usepackage{graphicx}
\usepackage{amsmath}
\usepackage{amssymb}
\usepackage{amsopn}
\usepackage{bm}

\begin{document}


\title{Optical cooling and trapping of
tripod-type atoms with rectified radiation forces.}


\author{I.~V.~Krasnov}
\email[]{krasn@icm.krasn.ru}
\affiliation{Institute of Computational Modeling,
Siberian Division, Russian Academy of
Sciences, 660036, Krasnoyarsk, Russia.
}


\date{\today}

\begin{abstract}
A new scheme of three-dimensional (\textit{3D}) all-optical (nonmagnetic) cooling and trapping of resonant atoms, based on using of so-called rectified radiation forces in non-monochromatic light fields is presented. It can be applied to the atoms with a tripod-type configuration of levels: atoms (ions) with the quantum transition $F=1\rightarrow F=0$.

 The  scheme proposed provides a long-term trapping of such atoms in deep light-induced potential wells. Moreover, the atom temperature can continuously be changed by varying field parameters in quite a large range (from super-Doppler to sub-Doppler values) without violating the localization stability.

\end{abstract}


\maketitle

\section{Introduction\label{}}
The problem of optical cooling and trapping of atoms plays one of the central roles in studying resonant light pressure. \cite{Kazantsev1990,Minogin1987,Metcalf2002}.
The most famous useful solutions of this problem are the \textit{3D} confinement and cooling of atoms in a magneto-optical trap (MOT) \cite{Raab1987} and optical molasses (OM) \cite{Chu1985}.
An indispensable element of MOT is a nonuniform magnetic field allowing one to ``circumvent'' the optical Earnshaw theorem (OET) \cite{Ashkin1983}.
In OM the principle of the \textit{3D} all-optical (nonmagnetic) viscous confinement of particles is manifested and, as contrasted to MOT, any confining space-dependent restoring force towards the center of OM is absent.
In OM the viscous damping force acts upon the atoms and the damping both of chaotic (resulting from laser cooling) and directed
(macroscopic) atom movement occurs. Therefore, the time of diffusion escape of the atoms from the OM area may be quite long as compared to the time of free expansion of the atom cloud. Note that OM is an example of an all-optical (nonmagnetic) device, allowing one to perform \textit{3D} cooling of the atoms to ultralow temperatures - below the so-called Doppler limit \cite{Metcalf2002}.
\par

The number of efficient all-optical methods of cooling and trapping of atoms can significantly be increased when using rectified radiation forces (RRFs), induced by non-monochromatic (particularly, biharmonic ) nonuniform fields such as standing waves. RRFs were discovered and described in \cite{Kazantsev1987,Kazantsev1989}, and experimentally demonstrated for the first time in \cite{Grimm1990}. Later, the idea of RRF was developed in many other studies, see, for example, \cite{Juvanainen1990,*Grimm1995,*Grimm1996,*Grove1995,*Pu1995,*Korsunsky1997,*Cashen2003} and the references given there.
\par

RRFs appear as the consequence of the nonlinear interference phenomena exhibited under the interaction of an atom with polychromatic electromagnetic field [8]. To explain their appearance physical mechanism we consider a simple model of the gradient force rectification, based on the results of the original works [7, 8]. Let a two-level atom moves with a velocity $v$ in a strong bichromatic standing wave field. Local Rabi frequency has the form
\begin{equation*}
\left( V_0(x)+V_1(x)e^{-i\Delta_1 t} \right)e^{-i\Delta_0 t},
\end{equation*}
where $V_0(x)=V_0\cos(kx)$, $V_1(x)=V_1\cos[(k+\delta k)x+\varphi]$, $\Delta_0$, $\Delta_1+\Delta_0$ are the detunings from resonance and we assume the following hierarchy of the characteristic frequencies ($\gamma$ is the spontaneous relaxation rate).
\begin{equation*}
\Delta_1\gg V_1\gg\Delta_0, \quad V_0, \quad \frac{V_1^2}{\Delta_1}\sim\frac{V_0^2}{\Delta_0}\gg kv, \quad \gamma.
\end{equation*}
Then, for weak saturation $V_1^2/\Delta_1\Delta_0$, $V_0^2/\Delta_0^2\ll 1$ and $\delta k=\Delta_1/c\ll k$, the mean dipole force $F$ (acting upon the atom) can be written as (more details see in [7, 8]):
\begin{equation*}
F=-\Pi_1\frac{dU_g}{dx}+\Pi_2\frac{dU_g}{dx}=Q\frac{dU_g}{dx},
\end{equation*}
where $\Pi_1$ and $\Pi_2$ are the populations of the adiabatic (dressed) atom states, $Q=\Pi_2-\Pi_1$ is the population difference, $U_g=V_0^2(x)/\Delta_0+V_1^2(x)/\Delta_1$ is potential of gradient force $F_g=-dU_g/dx$ determined by Stark (light) shifts of the atomic energy levels. This expression has a clear physical meaning: the atom in adiabatic state $\left|1\right>$ moves in the force field with the potential $U_g$ and atom in state $\left|2\right>$ moves in the force field with the potential $(-U_g)$. Spontaneous transitions cause the incoherent mixing of the adiabatic states, moreover the transition rates between the adiabatic states depend on the saturation parameter $V_0^2(x)/\Delta_0^2$ [1, 7, 8].

The result of this dependence is the spatial modulation (with the period $\sim\lambda/2=\pi/k$) of the population difference $Q=Q(x)$. In the adiabatic limit: $kv/\gamma\to 0$, $Q$ and the force $F$ can be presented as (see [1, 8])
\begin{equation*}
Q=-1+Q_1(x),\quad Q_1=\frac{2V_0^4(x)}{\Delta_0^4},
\end{equation*}
\begin{equation*}
F=-\frac{d\hat{U}_g}{dx} + F_{i}, \quad \hat{U}_g(x)=U_g(x)-\frac{2}{3}\frac{V_0^6(x)}{\Delta_0^5},
\end{equation*}
where the term $F_i=Q_1(x)dV_1^2(x)/dx$ one can interpret as the interference (nonadditive) contribution to the radiation force $F$ ($F_i=0$ if $V_1=0$ or $V_0=0$). Evidently, that the gradient force $\hat{F}_g=-(d\hat{U}_g/dx)$ does disappear (after averaging over the spatial oscillations with microscopic periods $\sim 1/k$), but the force $F_i$ does not disappear (as the change in sign of the gradient of the $V_1^2(x)$ can be compensated by the change in sign of the oscillating component of the $Q_1$ due to phase shift $\Psi$ between the two standing waves, $\Psi=\delta kx+\varphi$):
\begin{equation*}
\left< \hat{F}_g \right>_s= 0, \quad \left< F_i \right>_s=F_R=-\frac{dU_R}{dx},
\end{equation*}
\begin{equation*}
U_R=U_0\cos 2\Psi, \quad U_0=-\frac{\hbar}{4}\frac{k}{\delta k}\frac{V_0^4}{\Delta_0^4}\frac{V_1^2}{\Delta_1},\quad |U_0|\gg|\hat{U}_g|,
\end{equation*}
where $\left< \cdots \right>_s$ denotes the averaging over the microscopic spatial oscillations with the period of the order of light wavelength. The force $F_R$ is rectified gradient force by the terminology of [7, 8] as it has the order of magnitude of the gradient force, and it is constant-sign on macroscopic spatial scales $L\sim 1/\delta k$ much greater than the light wavelength $\lambda$.

Very similar physical scenarios of gradient force rectification where some frequency components of the optical field induce nonuniform redistribution of atoms over the quantum states and other frequency components form gradient forces (dependent on atomic state) can be organized for polychromatic fields of the \textit{3D} spatial configuration. In particular, this is demonstrated in the present paper. In this scenarios, the presence of the spatial microoscillations (of quantum state populations and gradient force potentials) with close frequencies or with phase shift each other is a necessary condition for the gradient force rectification.

So, the simple model described here and more general results of works [7-10] show that RRFs possess the following remarkable properties which allow one to use them to create \textit{3D} dissipative all-optical traps performing cooling and trapping of the particles. These forces are sign-constant at macroscopic spatial scales $L$ greatly exceeding the light wavelength $\lambda$ ($L\gg \lambda$), and in strong fields they have an order of magnitude of the induced light pressure force  (gradient force \cite{Kazantsev1990}) and are not saturated with increasing field intensity (thus, RRFs can significantly exceed the spontaneous light pressure force). The other useful property (which manifests both in strong and weak fields and which is not paid due attention to) is the controllability of the spatial structure of RRFs. In particular, choosing properly the configuration of the interfering light beams and their parameters it is possible to create the purely potential RRF (completely suppressing its vortex component). Such a RRF forms the system of deep potential wells \cite{Kazantsev1988} with a depth exceeding the characteristic value of the light (Stark) shifts of the atomic energy levels $\hbar \Delta_s$ by a large factor $L/\lambda\gg 1$.
\par

Moreover, in the general case, RRF contains a dissipative component --- friction force \cite{Kazantsev1989}.
\par

The RRFs properties mentioned guarantee the possibility of complete overcoming OET constraints and reveal the perspective of their use to create new all-optical schemes of cooling and trapping of atoms.
\par

However, certain solutions significantly depend on the structure of atomic transitions. The \textit{3D} atom localization $^{85}\mathrm{Rb}$
in an optical superlattice (induced by RRF) was demonstrated by experiment in \cite{Gorlitz2001}. This paper notes an interesting possibility of applying such methods of the atom localization in quantum computing operations. In \cite{Wasik1997} the authors investigate the possibility of using RRFs for \textit{3D} all-optical trapping and cooling of the atoms with the quantum transition $F_b=1/2 \rightarrow F_a=3/2$ (where $F_b$ and $F_a$ are the total angular momenta in the ground and excited states, respectively).
\par

Based on the performed \textit{3D} semiclassical Monte-Carlo simulations of the atom motion in the biharmonic field (\textit{3D} configuration), it was shown that in the case considered the stable atom localization in the \textit{3D} superlattice can be combined with their sub-Doppler cooling.
\par

In \cite{Krasnov2002, Krasnov2004, Krasnov2001} the three-dimensional rectification of a radiation force in weak and strong non-monochromatic fields in the case of the atoms with the quantum transition $F_b=0\rightarrow F_a=1$ was investigated, certain all-optical schemes of the \textit{3D} stable confinement of such atoms were suggested. The temperature of the trapped atoms in the cases considered in \cite{Krasnov2002,Krasnov2004,Krasnov2001} can not exceed the value of the order of Doppler cooling limit.
\par

In the present work the problem of all-optical \textit{3D} trapping and cooling of resonant particles is studied using RRFs for the cases of the atoms (ions) with the tripod-type configuration of the working levels: for the atoms (ions) with the quantum transition $F_b=1 \rightarrow F_a=0$. The case considered is significantly different from that of the atoms with the quantum transition  $F_b=0 \rightarrow F_a=1$ due to the degeneracy of the ground state. This important factor predetermines the possibility of the sub-Doppler cooling. On the other hand (as we shall see later) the tripod-type atom is convenient for the theoretical analysis of the {\it 3D} problems.
\par

It was shown that RRFs, induced by non-monochromatic optical fields are able to provide the stable deep \textit{3D} localization of tripod-type atomic particles and efficient control of their temperature.
\par

In the proposed \textit{3D} dissipative optical trap the temperature of the localized atoms can be varied in a very wide range of values including both super-Doppler and sub-Doppler temperatures.
\par

The desired effect (trapping and cooling of atoms) can be achieved using the \textit{3D} scheme for rectifying a gradient force based on employing the partially coherent optical field involving (besides the  coherent components) components with fluctuating phases (the same idea for the case of atoms with the quantum transition $F_b=0 \rightarrow F_a=1$ was considered in \cite{Krasnov2004})
\par

A distinctive peculiarity of the investigation conducted is the approximate analytical description of the \textit{3D} effects of the RRF action on the tripod-type atoms in terms of the kinetic quasi-classical theory of the radiation forces based on the Wigner density matrix formalism, allowing one to correctly take into consideration the quantum fluctuations of the radiation forces. Note that in most works devoted to RRFs (see \cite{Juvanainen1990,*Grimm1995,*Grimm1996,*Grove1995,*Pu1995,*Korsunsky1997,*Cashen2003}) the  \textit{1D} problems are usually analyzed and RRFs are determined using the conventional optical Bloch equations (the exception from the papers cited is the article \cite{Krasnov2002}).
\par

The paper is organized as follows. Section II describes the kinetic quasi-classical model of the mechanical action of non-monochromatic light on the  tripod-type atoms and procedure of averaging equations for the Wigner density matrix, resulting in their significant simplification.
\par

Section III describes in detail the \textit{3D} configuration of the optical fields which provides the desired effect of the stable localization and cooling of atoms.
\par

In Section IV the Fokker-Plank equation for the Wigner distribution function of the atoms  $f(\mathbf{r},\mathbf{v},t)$ is obtained, and based on the analysis  of its coefficients from the governing parameters different characteristic regimes of the mechanical action of optical fields on the tripod-type atoms are described.
\par

In Section V an asymptotic velocity distribution of the atoms is found and the Smoluchowsky equation for the configuration space distribution function $n=n(\mathbf{r},t)$, describing the slow (diffusion) stage of the evolution of an atom ensemble is obtained. The possibility of the long-term spatial localization of the  atoms in deep potential wells is shown and the dependence of temperature of the trapped atoms on the field parameters is analyzed. Particularly, the possibility of the stable long-term trapping of atoms both with Super-Doppler and Sub-Doppler temperatures is demonstrated.

\section{Model\label{}}

Consider an ensemble of atoms in the light field
\[\mathbf{E}(\mathbf{r},t)e^{-i\omega_0 t}+c.c.,\]
with the carrier frequency $\omega_0$, tuned resonant to $\lvert F_b=1, M_b=0, \pm 1 \rangle \rightarrow \lvert F_a=0, M_a=0 \rangle$ closed (cycling) atomic transition, where $F_\alpha$, $M_\alpha$  is the full angular momentum and its projections for the ground $\alpha=b$ and excited  $\alpha=a$ states. The field is a superposition of the coherent quasi-resonant components (with three different frequencies), polarized in mutually perpendicular directions and partially coherent (fluctuating) resonant field $\mathbf{E}'$ with the bandwidth $\sim\Gamma$:
\begin{equation}\label{f1}
\mathbf{E}\left(\mathbf{r},t\right) = \sum_{j=x,y,z} E_{j1}(\mathbf{r})\mathbf{e}_j\exp\lbrack -i\Delta_j t \rbrack + \mathbf{E}'(\mathbf{r},t),
\end{equation}
where $\mathbf{e}_j$ denotes the unit basis vectors of the Cartesian coordinate system and $\Delta_j$ is detuning from the resonant frequency $\omega_0$. According to the original conception of the effect of the radiation force rectification \cite{Kazantsev1987, Kazantsev1989} (compare also with \cite{Krasnov2004}) one can assume the following hierarchy of the characteristic frequencies:
\[ \Delta_j, \quad \lvert \Delta_j-\Delta_l \rvert \gg \lvert V_{j1} \rvert, \]
\begin{equation}\label{f2}
\Gamma\gg \lvert U_j \rvert ,\quad\frac{\lvert V_{j1}\rvert^2}{\Delta_j}, \quad\frac{\lvert U_j\rvert^2}{\Gamma},\quad\gamma, \quad ks,
\end{equation}
\begin{equation}\label{f3}
\gamma\left\lvert\frac{V_{j1}}{\Delta_j}\right\rvert^2\ll\frac{\lvert U_j\rvert^2}{\Gamma},
\end{equation}
where $l$ and $j\neq l$ denote the indices $x$, $y$ or $z$, $V_{j1}=dE_{j1}^*/\hbar$, $U_j=d(\mathbf{e}_j\cdot\mathbf{E}'^*)/\hbar$ are the Rabi frequencies,
$d=\lVert d\rVert/\sqrt{3}$, $\lVert d\rVert$ is the reduced dipole transition matrix element, $k=\omega_0/c$ is the wave number, $s$ is the thermal velocity of atoms (characteristic width of the velocity distribution of atoms), $\gamma = \gamma'/3$, $\gamma'$ is the rate of the spontaneous decay of the excited state. The relations between the frequencies in the right part of inequality (\ref{f2}) can be arbitrary. Inequalities (\ref{f2})--(\ref{f3}) imply that the coherent components of the field are ``quasi-resonant'', and the fluctuating component $\mathbf{E}'$ is ``resonant''. Therefore, (as it will be seen later) the coherent components of the field, $\propto E_{j1}(\mathbf{r})\mathbf{e}_j$, form the spatially non-uniform light (Stark) shifts of the atomic energy levels, and the fluctuating field component $\mathbf{E}'(\mathbf{r},t)$ provides incoherent excitation of the atoms (their redistribution  over the quantum states). In the scheme of the gradient force rectification in the bichromatic field considered in \cite{Kazantsev1987, Kazantsev1989} this effect is achieved due to the presence of the coherent field component with a relatively low frequency detuning.
\par

The state of the atoms interacting with the optical field will be described using the Wigner density matrix $\hat{\rho}(\mathbf{r},\mathbf{v},t)$ \cite{Kazantsev1990,Minogin1987}. In the quasi-classical limit, when the photon momentum is much lower than the characteristic atomic momentum distribution width, $\hbar k\ll ms$ (where $m$ is the atomic mass), and in the interaction representation, this density matrix satisfies the quantum kinetic equation \cite{Kazantsev1978,Rautian1979}
\begin{equation}\label{f4}
\frac{d\hat{\rho}}{dt}+\hat{\gamma}\hat{\rho}=-i\left[\hat{V}\hat{\rho}\right]+\frac{1}{2m}\left\{ \frac{\partial\hat{V}}{\partial\mathbf{r}} \frac{\partial\hat{\rho}}{\partial\mathbf{v}} \right\},
\end{equation}
\[ \frac{d}{dt}=\frac{\partial}{\partial t}+\mathbf{v}\cdot\frac{\partial}{\partial\mathbf{r}}, \]
where $\hbar\hat{V}$ is the dipole atom-field interaction Hamiltonian, $\hat{\gamma}$ is the relaxation operator that includes the recoil effect during spontaneous transition \cite{Kazantsev1990,Minogin1987}, and the square brackets and braces denote the commutator and anticommutator, respectively. The second term in the right part of Eq.(\ref{f4}) takes into account the recoil effect in induced transitions. Further, it is convenient for our analysis to consider $\hat{\rho}$ in the Cartesian representation (compare with \cite{Kazantsev1985,Krasnov2002,Krasnov2004}) i.e. in the representation of the basis wave functions (of intra-atomic motion) for the excited $|a\rangle$ and ground states $|bi\rangle$:
\[ |a\rangle=|0,0\rangle, \quad |bz\rangle=|1,0\rangle, \]
\[ |bx\rangle=\frac{|1,-1\rangle-|1,1\rangle}{\sqrt{2}}, \quad |by\rangle=i\frac{|1,-1\rangle+|1,1\rangle}{\sqrt{2}},\]

In this representation the matrix elements of the dipole moment $\hat{\mathbf{d}}$ are directed along the unit vectors of the Cartesian coordinate system,
\begin{equation*}
\langle bi|\hat{\mathbf{d}}|a\rangle=\mathbf{e}_i d, \quad i=x,y,z,
\end{equation*}
and the Hamiltonian $\hbar\hat{V}$ in the rotating-wave approximation ($|\Delta_j|$, $\Gamma\ll\omega_0$) has the following form:
\begin{eqnarray}\label{f5}
\hbar\hat{V}&=&-\hbar\sum_{j=x,y,z} [ ( V_{j1}(\mathbf{r})\exp(i\Delta_j t) \nonumber \\
&&+U_j(\mathbf{r},t) ) |b_j\rangle \langle a| + H.c. ]
\end{eqnarray}
\par

For the density matrix elements in the Cartesian representation the following notations will be used:
$ \langle bi|\hat{\rho}|bi\rangle=\rho_{ii}(\mathbf{r},\mathbf{v},t) $,
$ \langle a|\hat{\rho}|a\rangle=\rho(\mathbf{r},\mathbf{v},t) $,
$ \langle bi|\hat{\rho}|bj\rangle=q_{ij}(\mathbf{r},\mathbf{v},t) $
with $i\neq j$, $ \langle bi|\hat{\rho}|a\rangle=\rho_i(\mathbf{r},\mathbf{v},t)$. Thus, $\rho_{ii}$, $\rho$  are the Wigner distribution functions for the atoms  in the states $|bi\rangle$ and $|a\rangle$ , respectively, the functions $\rho_i$ determine the projections of the induced dipole moment onto the axes of the Cartesian coordinate system, $q_{ij}$ characterizes the coherence between $|bi\rangle$ and $|a\rangle$ states.
\par

From Eq.(\ref{f5}) it follows that the coefficients of the equations for the density matrix elements (see Eq.(\ref{f4})) oscillate with the frequencies $\Delta_j(j=x,y,z)$ and, moreover, contain fluctuating (stochastic) components. Taking into account inequalities (\ref{f2}),(\ref{f3}) we average successively Eq.(\ref{f4}) --- at first, over the high-frequency oscillations with the frequencies $\Delta_j$ (as is usually done in the RRFs theory (see \cite{Kazantsev1987,Kazantsev1989,Krasnov1994}), then, over the fluctuations of the field $\mathbf{E}'$. At the first stage one obtains the system of equations (using the same notation for the averaged quantities):
\begin{equation}\label{f6}
i\left( \frac{d}{dt} + \gamma_{\perp} -i\hat{\Delta}_i(\mathbf{r}) \right) \rho_i = \sum_{j=x,y,z}q_{ij}U_j+\Lambda_i,
\end{equation}
\begin{widetext}
\begin{equation}\label{f7}
i\left( \frac{d}{dt} - i\hat{\Delta}_{ij}(\mathbf{r}) \right) q_{ij} + i\gamma\delta_{ij}\sum_{l=x,y,z}q_{ll}=
i\gamma f\delta_{ij}+(\rho_i U_j^*-U_i\rho_j^*) + \delta_{ij}\sum_{l=x,y,z}(\rho_l U_l^*-c.c.)+B_{ij},
\end{equation}
\begin{equation}\label{f8}
\frac{df}{dt}+\frac{\hbar}{m} \sum_{i=x,y,z} \left( \frac{\partial\rho_i}{\partial\mathbf{v}}\cdot\nabla U_i^*+c.c. \right)
-\frac{\hbar}{m}\sum_{i=x,y,z}\frac{\partial q_{ii}}{\partial\mathbf{v}}\cdot\frac{\nabla|V_{i1}|^2}{\Delta_i}
=\hat{D}_s \left( f-\sum_{i=x,y,z}q_{ii}, \right)
\end{equation}
\end{widetext}
where $f(\mathbf{r},\mathbf{v},t)={\rm Sp}(\hat{\rho})=\rho+\sum_i \rho_{ii}$ is the Wigner particle distribution function (DF) in the phase space $(\mathbf{r},\mathbf{v})$, $q_{ii}=\rho_{ii}-\rho$ denotes the densities of the distribution of the population difference,
\[ \hat{\Delta}_i(\mathbf{r}) = -\left[ \frac{2|V_{i1}(\mathbf{r})|^2}{\Delta_i} + \sum_{l\neq i} \frac{|V_{l1}(\mathbf{r})|^2}{\Delta_l}\right] \]
are the effective spatially nonuniform (in case of the nonhomogeneous fields) detunings due to the light-induced Stark shifts, $\hat{\Delta}_{ij}(\mathbf{r})=\hat{\Delta}_i(\mathbf{r})-\hat{\Delta}_j(\mathbf{r})$,
\begin{equation}\label{f9}
 \hat{D}_s = \left( \frac{\hbar k}{m}\ \right)^2 \frac{\gamma}{8} \left( \frac{\partial}{\partial\mathbf{v}} \right)^2
\end{equation}
is the operator describing the recoil effect at spontaneous transitions \footnote{This expression results from the general expression for the relaxation operator $\hat{\gamma}$ given in the book \cite{Kazantsev1990} when passing to the Cartesian representation for $\hat{\rho}$}. The terms $\Lambda_i$ and $B_{ij}$ in the right part of Eqs. (\ref{f6}), (\ref{f7}) describe the influence of the recoil effect on the induced transitions and coherence between  $|bi\rangle$ and $|a\rangle$ states. Explicit expressions for them are given in Appendix, as well as some comments on the effects described in Eqs.(\ref{f6})--(\ref{f8}).
\par

Next, Eqs. (\ref{f6}), (\ref{f7}) are averaged over the fluctuations of the field $\mathbf{E}'$ \footnote{Note, that the radiation forces due to the fluctuating fields with a finite bandwidth were originally considered in \cite{Cook1980} for the case of two-level atoms (TLA). A little earlier (also for the case of TLA) the averaged Fokker-Planck equation describing atom heating in the field of the standing wave with the fluctuating phase was obtained (see review \cite{Kazantsev1978} and references there).}. These equations are the system of multiplicative stochastic linear equations and the standard procedure described in \cite{Kampen1984} was used to average them. In the problem under consideration it is based on the expansion of the solution of Eqs. (\ref{f6}), (\ref{f7}), (\ref{f8}) in terms of $\zeta\ll 1$, which is proportional to the correlation time $\tau_c\sim\Gamma^{-1}$:
\[|U_j|\tau_c,\;|\hat{\Delta}_j|\tau_c,\; ks\tau_c,\; \gamma\tau_c\le\zeta\ll 1.\]
The smallness of the parameter $\zeta$ is due to the original assumptions: the right part of inequality (\ref{f2}), i.e. due to the conditions which determine the value of the bandwidth $\Gamma$ of the field $\hat{\mathbf{E}}'$. Moreover, one makes the following additional assumptions on the properties of the random processes $U_j,\; j=x,y,z$:
\begin{equation}\label{f10}
\langle\!\langle U_j\rangle\!\rangle=0,
\langle\!\langle U_j(\mathbf{r},t)U_i(\mathbf{r}',t+\tau)\rangle\!\rangle=C_i(|\tau|,\mathbf{r},\mathbf{r}')\delta_{ji},
\end{equation}
where the double angular brackets denote averaging over the fluctuations.
\par

Thus, suppose that $U_j$ is treated as a stationary random process with zero-mean, and the $\mathbf{E}'$ components with different polarization fluctuate independently. As a result, the following closed system of the reduced equations describing the kinetics of the tripod-type atoms in the light field was obtained (without changing notations for the averaged DF):
\begin{eqnarray}\label{f11}
\left[ \frac{d}{dt}+2R_i(\mathbf{r})+\gamma \right]q_i &+&\sum_{j\ne i}[R_j(\mathbf{r})+\gamma]q_j  \nonumber \\
&=&\gamma f+B_i,\quad i=x,y,z,
\end{eqnarray}
\begin{eqnarray}\label{f12}
 \frac{df}{dt}&+&\sum_{i}\frac{\partial}{\partial\mathbf{v}m} \left( \mathbf{F}_i-\nabla\frac{\hbar|V_{i1}|^2}{\Delta_i} \right)q_i   \nonumber \\
&=&\hat{D}_s\left( f-\sum_{j}q_j \right)+\hat{S}\{Q\},
\end{eqnarray}
\vspace{1pt}
\begin{equation}\label{f13}
\hat{S}\{Q\}=\sum_{i,j,j'}\frac{\partial^2Q_i}{\partial v_j\partial v_{j'}}A_{jj'}^i,
\end{equation}
where $q_i=\langle\!\langle q_{ii} \rangle\!\rangle$, $Q_i=\langle\!\langle Q_{ii} \rangle\!\rangle$, $Q_{ii}$ is determined by the expression (\ref{a1}), $v_j=\mathbf{v}\cdot\mathbf{e}_j$, and $R_i$, $\mathbf{F}_i$, $A_{jj'}^i$ are determined by the correlators
\begin{equation}\label{f14}
R_i(\mathbf{r})=2 \Re e\int\limits_{-\infty}^0 \langle\!\langle U_i(\mathbf{r},t)U_i^*(\mathbf{r},t+\tau) \rangle\!\rangle d\tau,
\end{equation}
\begin{equation}\label{f15}
\mathbf{F}_i=-2\hbar \Im m\int\limits_{-\infty}^0 \langle\!\langle \nabla U_i(\mathbf{r},t)U_i^*(\mathbf{r},t+\tau) \rangle\!\rangle d\tau,
\end{equation}
\begin{eqnarray}\label{f16}
A_{jj'}^i & = & \frac{\hbar^2}{2m^2} \Re e\int\limits_{-\infty}^0
\left\langle\!\!\!\left\langle
    \frac{\partial U_i^*(\mathbf{r},t)}{\partial r_j}
    \frac{\partial U_i(\mathbf{r},t+\tau)}{\partial r_{j'}}
\right\rangle\!\!\!\right\rangle d\tau, \nonumber \\
\frac{\partial}{\partial r_j} & = & (\mathbf{e}_j\cdot\nabla).
\end{eqnarray}
The  coefficients $R_i(\mathbf{r})$ have the meaning of the rates of the transitions between the low-lying $|bi\rangle$ and excited $|a\rangle$ atomic states induced by the field $\mathbf{E}'$, the value $\mathbf{F}_i$ has the dimension of force and is proportional to the energy flux density of the $\mathbf{E}'$ field component, polarized along $\mathbf{e}_i$, the coefficients $A_{jj'}^i$ have the dimension of velocity diffusion coefficient and the order of magnitude $(\hbar k/m)^2R_i$. The last term in the right part of equation (\ref{f11}) $B_i=-i\langle\!\langle B_{ii} \rangle\!\rangle$ (where $B_{ii}$ is determined by Eq. (\ref{a4}) and has the following  explicit representation
\begin{eqnarray}\label{f17}
B_i & = & \frac{1}{4m}\sum_{j\ne i}\frac{\partial}{\partial\mathbf{v}}
\left[ (Q_j-2q_j)\nabla\frac{\hbar|V_{j1}|^2}{\Delta_j} +  (Q_j+2q_j)\mathbf{F}_j \right]   \nonumber \\
&& + \frac{1}{2m}\frac{\partial Q_i}{\partial \mathbf{v}}\cdot
\left[ \nabla\frac{\hbar|V_{i1}|^2}{\Delta_i}+\mathbf{F}_i \right].
\end{eqnarray}
\par

The system of four equations (\ref{f11})--(\ref{f12}) is much more simple than the system of sixteen ones (\ref{f6})--(\ref{f8}) for the elements of the density matrix. Such a simplification results from the original assumptions on the relations between the parameters of the optical field and the properties of the field component correlators: i.e. from inequalities (\ref{f2}) и (\ref{f3}) and Eqs. (\ref{f10}). Following the generally accepted approach of the quasi-classical theory of the mechanical action of light on atoms \cite{Kazantsev1990,Minogin1987}, later (in the Section IV) one obtains from Eqs. (\ref{f11}), (\ref{f12}) the Fokker-Planck equation (FPE) for DF $f(\mathbf{r},\mathbf{v},t)$; however, to obtain the explicit FPE representation it is necessary to specify the spatial configuration of the optical fields. In the next Section, in accordance with the aims of this work, such a field configuration will be taken where the \emph{3D} kinetics of the atoms is completely determined by RRFs.

\section{Optical field configuration \label{}}
Let the optical field be formed by a special superposition of the plane light waves, where the Rabi frequencies $V_{i1}$ and $U_i$ ($i=x,y,z$) are the following:
\begin{equation}\label{f18}
V_{i1}(\mathbf{r})=\frac{V_i}{2}\left[ \exp(i\mathbf{q}_i\cdot\mathbf{r}+i\eta_1) + \exp(i\mathbf{q}_i'\mathbf{r}) \right],
\end{equation}
\begin{equation}\label{f19}
U_i(\mathbf{r})=\sum_{\alpha=1}^4 U_{i\alpha}(\mathbf{r})\exp\left(i\varphi_\alpha(t)\right),
\end{equation}
\begin{eqnarray}\label{f21}
U_{i1}(\mathbf{r})&=&\frac{U}{2}\left[ \exp(i\mathbf{k}_{i1}\cdot\mathbf{r}) + a\exp(i\mathbf{k}_{i1}'\cdot\mathbf{r}) \right], \nonumber \\
U_{i3}(\mathbf{r})&=&\frac{\sqrt{a'}U}{2}\left[ \exp(i\mathbf{k}_{i3}\cdot\mathbf{r}+i\zeta_1)+\exp(i\mathbf{k}_{i3}'\cdot\mathbf{r}) \right], \nonumber \\
U_{i2}(\mathbf{r})&=&U_{i1}^*(\mathbf{r}),\quad U_{i4}(\mathbf{r})=U_{i3}^*(\mathbf{r}),
\end{eqnarray}
where the index $i=x,y,z$, $\eta_1$, and $\zeta_1$ are the fixed phase shifts, the positive parameters $a$, $a'\ll1$, $U$ and $V_i$ are the real amplitudes (which in the considered model do not depend on $\mathbf{r}$), $\varphi_{i\alpha}(t)$ are independently fluctuating phases (with delta-correlated zero-mean derivatives), which determine the correlators of the $\mathbf{E}'$ components by the relations
\begin{equation}\label{f22}
\langle\!\langle \exp i(\varphi_{j\alpha}(t)-\varphi_{l\beta}(t+\tau)) \rangle\!\rangle = \delta_{jl}\delta_{\alpha\beta}\exp(-\Gamma|\tau|).
\end{equation}
\par

Thus, the field $\mathbf{E}'$ is described by the so-called phase-diffusion model \cite{Kazantsev1978, Cook1980, Stenholm1984} and has a Lorentzian spectral profile $J(\omega)$ with bandwidth $\Gamma$:
\begin{equation*}
J(\omega)\propto\frac{\Gamma}{(\omega-\omega_0)^2+\Gamma^2}.
\end{equation*}
Here, all the assumptions on the properties of the random processes $U_j$ are implemented (see Eqs.(\ref{f10})).
\par

It is worth noting that the representation (\ref{f19}) for $U_i(\mathbf{r},t)$ is valid only if the coherence length $l_c=c\tau_c=c/\Gamma$ is much greater than the characteristic atomic cloud size $D$ (see detailed discussion in \cite{Cook1980})
\begin{equation}\label{f23}
l_c\gg D.
\end{equation}
\par

Specifying the configuration of the optical field, one determines the wave vectors in Eqs.(\ref{f18}) as follows (see Fig.1):
\begin{figure}
\includegraphics{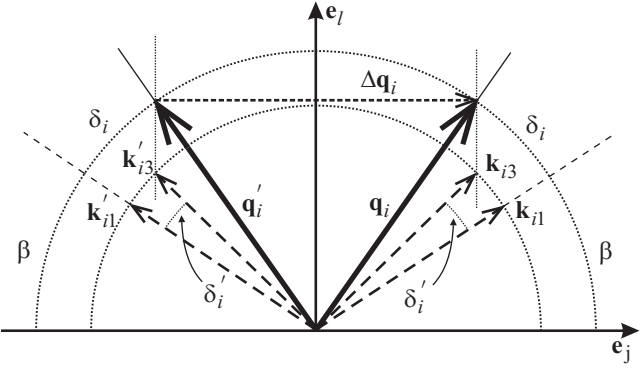}
\caption{
The wave vectors of the light field components polarized along $\mathbf{e}_i$ and determining the Rabi frequencies in Eqs. (\ref{f18}--\ref{f21}), the ordered index triplets $(ijl)$ are: $(xyz)$, $(zxy)$ and $(yzx)$, $\delta_i$ and $\delta_i'$ being small angular detunings. The bold arrows are the wave vectors corresponding to the coherent field components. The vectors $\mathbf{k}_{i2}=-\mathbf{k}_{i1}$, $\mathbf{k}_{i2}'=-\mathbf{k}_{i1}'$, $\mathbf{k}_{i4}=-\mathbf{k}_{i3}$, $\mathbf{k}_{i4}'=-\mathbf{k}_{i3}'$ determining $U_{i2}(\mathbf{r})$ and $U_{i4}(\mathbf{r})$ are not presented in the Figure. Thus, each component $E_i'\mathbf{e}_i$ of the field $\mathbf{E}'$ is a sum of four pairs of the counter-propagating waves.
\label{}}
\end{figure}
\begin{eqnarray}\label{f24}
\mathbf{k}_{i1}&=&k(\mathbf{e}_j \cos\beta+\mathbf{e}_l\sin\beta), \nonumber \\
\mathbf{k}_{i1}'&=&k(-\mathbf{e}_j \cos\beta+\mathbf{e}_l\sin\beta), \nonumber \\
\mathbf{q}_i&=&q_i(\mathbf{e}_j\cos\beta_i + \mathbf{e}_l\sin\beta_i), \nonumber \\
\mathbf{q}_i'&=&q_i( -\mathbf{e}_j\cos\beta_i+\mathbf{e}_l\sin\beta_i ), \nonumber \\
\mathbf{k}_{i3}&=&k(\mathbf{e}_j \cos\beta_i'+\mathbf{e}_l'\sin\beta_i'), \nonumber \\
\mathbf{k}_{i3}'&=&k(-\mathbf{e}_j \cos\beta_i'+\mathbf{e}_l'\sin\beta_i'),
\end{eqnarray}
where $\beta$ is an angle determining the dominant directions of the propagation of the waves polarized along $\mathbf{e}_i$, $\beta_i=\beta+\delta_i$, $\beta_i'=\beta+\delta_i'$, $\delta_i$ and $\delta_i'$ are small angular detunings ($|\delta_i|,|\delta_i'|\ll 1$), and the ordered index combinations $(ijl)$ are: $(xyz)$, $(zxy)$, $(yzx)$. The angular detunings $\delta_i$ and $\delta_i'$ are free parameters allowing one to control the macroscopic spatial structure of the field (compare with \cite{Kazantsev1989}), we choose them in such a way that the following conditions are satisfied for the difference of the wave vectors:
\begin{equation}\label{f25}
\Delta \mathbf{q}_i=\mathbf{q}_i-\mathbf{q}_i'=\mathbf{k}_{i3}-\mathbf{k}_{i3}'=\mathbf{k}_{i4}'-\mathbf{k}_{i4},
\end{equation}
\begin{equation*}
\Delta \mathbf{K}_i=\Delta\mathbf{q}_i-\Delta\mathbf{k}_i = 2k\alpha_i\xi_i\cos\beta\mathbf{e}_j=\frac{2\pi}{L_i}{\rm sgn}(\xi_i\alpha_i)\cos\beta\mathbf{e}_j,
\end{equation*}
where $\alpha_i=(\Delta_i/\omega_0)\ll1$, $\Delta\mathbf{k}_i=\mathbf{k}_{i1}-\mathbf{k}_{i1}'=\mathbf{k}_{i2}'-\mathbf{k}_{i2}$, the vectors $\mathbf{k}_{i2}=-\mathbf{k}_{i1}$, $\mathbf{k}_{i2}'=-\mathbf{k}_{i1}'$, $\mathbf{k}_{i4}=-\mathbf{k}_{i3}$, $\mathbf{k}_{i4}'=-\mathbf{k}_{i3}'$ determine $U_{i2}(\mathbf{r})$ and $U_{i4}(\mathbf{r})$ in the superposition (\ref{f19}) (in accordance with Eqs. (\ref{f21})), the parameters $\xi_i=(1-\delta_i^2/2\alpha_i-\delta_i\tan\beta)$ determine the macroscopic spatial scales of the problem $L_i$ (i.e. as one will see, the period of the spatial RRFs modulation): $L_i=\pi c/|\Delta_i\xi_i|\gg\lambda=2\pi/k$. Further, we will restrict ourselves to the case of the configurations of the optical fields and values of the angles $\beta$ close to $\pi/2$ or $0$, for which the following relations hold
\begin{equation}\label{f26}
L_i=L,\quad i=x,y,z,\quad\beta_1=|\sin2\beta|\ll 1.
\end{equation}
\par

Then, given the values $\Delta_i$ and $L\gg\lambda$ and not very small values of $\beta$ ($\tan\beta\gg\sqrt{|\alpha_i\xi_i|}$), the necessary (to fulfill Eq.(\ref{f25}) and the first one from Eqs.(\ref{f26})) values of the governing parameters $\delta_i$ and $\delta_i'$ are described by the simple formulae:
\begin{equation}\label{f27}
\delta_i\simeq\frac{(1-\xi_i)\alpha_i}{\tan\beta},\quad \delta_i'\simeq-\frac{\alpha_i \xi_i}{\tan\beta},
\end{equation}
where $|\xi_i|=\pi c/|\Delta_i|L$, and the sign $\xi_i$ may be taken arbitrarily.

So the following conclusions can be drawn: each group of light waves polarized along $i$-axis can produce an interference pattern along $j$-axis (where index $j=y$ for $i=x$; $j=x$ for $i=z$; $j=z$ for $i=y$). Spatial frequencies of this interference patterns $\Delta \mathbf{q}_i$ and $\Delta\mathbf{k}_i$ are close to each other (due to smallness of angular detunings $\delta_i$, $\delta_i'$) and determine microscopic spatial scale of the problem $\lambda_M\simeq\lambda/2\cos\beta\sim 1/|\Delta\mathbf{k}_i|,1/|\Delta\mathbf{q}_i|$ while their difference $\Delta\mathbf{K}_i$ determines macroscopic spatial scale:
$L\sim(1/|\Delta\mathbf{K}_i|)\gg\lambda_M$.

In the field configuration considered the explicit expressions for the transition rates $R_i(\mathbf{r})$ and the term $\hat{S}(Q)$ in the right part of equation (\ref{f12}) for DF $f(\mathbf{r},\mathbf{v},t)$ are the following
\begin{equation*}
R_i(\mathbf{r})=R(1+a_1 P_i(\mathbf{r})),\quad R=\frac{U^2}{\Gamma}(1+a^2+2a'),
\end{equation*}
\begin{equation}\label{f28}
P_i(\mathbf{r})=\frac{1}{1+b}\left[ \cos(\Delta\mathbf{k}_i\cdot\mathbf{r})+b\cos(\Delta\mathbf{q}_i\cdot\mathbf{r}+\xi_1) \right],
\end{equation}
\begin{equation}\label{f29}
\hat{S}\{Q\}\approx\frac{\hbar^2k^2}{4m^2}R\sum_{i=x,y,z}\left( \cos^2\beta\frac{\partial^2}{\partial v_j^2} + \sin^2\beta\frac{\partial}{\partial v_l^2} \right)Q_i,
\end{equation}
where $a_1=2(a+a')/(1+a^2+a')$, $b=a'/a$ and in Eq.(\ref{f29}) the ordered index triplets $(ijl)$ are: $(xyz)$, ($zxy$), ($yzx$). The most important peculiarity of the field configuration determined by Eqs.(\ref{f18})--(\ref{f21}) is vanishing of the forces of the resonant light pressure, conditioned only by the fluctuating field $\mathbf{E}'$. This results from the accurate mutual compensation of independent (due to  Eqs.(\ref{f22})) contributions into $\mathbf{F}_i$ (see Eqs.(\ref{f15}), proportional to the energy flux density of the field components with the Rabi frequencies $U_{i\alpha}$, $\alpha=1-4$:
\begin{equation}\label{f30}
\mathbf{F}_i=0.
\end{equation}
\par

Eqs. (\ref{f30}) hold in case when Eqs. (\ref{f21})are satisfied.
\par

Now one can clearly see the final distribution of the ''roles`` of the coherent $\mathbf{E}_1$ and partially coherent $\mathbf{E}'$ fields in the considered model of the mechanical action of light on the tripod-type atoms, see Eqs. (\ref{f11}), (\ref{f12}). The fluctuating field $\mathbf{E}'$ is responsible for incoherent mixing of the atomic states (or in other words – for the redistribution of the atoms over the quantum states), and the quasi-resonant coherent field induces the effective potentials (due to the light-induced Stark shift of the energy levels), which determine the motion of the atoms: the unexcited atoms (in the state $|bi\rangle$) move in the field of the gradient force with the potential (oscillating  with a period of the order of the light wave length $\lambda$)
\begin{equation*}
\frac{\hbar|V_{i1}\mathbf{r}|^2}{\Delta_i}, \quad i=x,y,z,
\end{equation*}
the excited atoms (in the state $|a\rangle$) in the field of the gradient force with the potential
\begin{equation*}
-\sum_{i=x,y,z}\frac{\hbar|V_{i1}(\mathbf{r})|^2}{\Delta_i}.
\end{equation*}

In order to better understand the above-described physical picture, it is sufficient to pay attention on the fact that Eqs.\,(11),\,(12) (at condition Eq.\,(29)) are equivalent to the following equations for Wigner distribution functions $\rho_{ii}$ and $\rho$ for the atoms in the states $\left| bi\right>$ $(i=x,y,z)$ and $\left| a\right>$ ($\rho=(f-\sum q_{ii})/4$, $\rho_{ii}=\rho+q_{ii}$)
\begin{equation*}
\frac{d\rho_{ii}}{dt}-\frac{\nabla\hbar|V_{i1}|^2}{\Delta_i}\frac{\partial\rho_{ii}}{\partial\mathbf{v}}=\gamma\rho-R_i(\mathbf{r})(\rho_{ii}-\rho)+S_i,\quad i=x,y,z,
\end{equation*}
\begin{equation*}
\frac{d\rho}{dt}+\left( \sum_i\frac{\nabla\hbar|V_{i1}|^2}{\Delta_i} \right)\frac{\partial\rho}{\partial\mathbf{v}}=-3\gamma\rho+\sum_i R_i(\mathbf{r})(\rho_{ii}-\rho)+S,
\end{equation*}
where $S_i$ and $S$ are second-order terms in the quasi-classicality parameter $\hbar k/ms$  responsible for the velocity diffusion (conditional on spontaneous transitions and fluctuating field $\mathbf{E}'$ influence). These equations clearly demonstrate four-potential nature of atomic kinetics and a factor of the incoherent redistribution of population of quantum states (relevant to these four potentials). The mean gradient force is obviously, a sum of gradient forces weighted by the  probabilities of occupation $\Pi_i=\rho_i/f$ $(i=x,y,z)$,  $\Pi=\rho/f$ of atom states.

Therefore, the physical model proposed is a generalization of the simple model of the mechanical action of bichromatic standing wave on two-level atom, which is described in the Introduction. The only difference is the number of atomic states, more complicated spatial structure of gradient force potentials and much more complex nature of atomic population redistribution among quantum states.

Taking into consideration this analogy it is possible to assume that the necessary condition of the gradient force rectification in our problem (cf [15]) is the spatial modulation (with the period $\sim\lambda$) of the transition rates $R_i(\mathbf{r})$, and hence the spatial modulation of the relative population differences $\hat{q}_i=q_i/f$, which is possible only if the field $E'$ has mutually interfering components, i.e. when $a, a'\ne 0$. Though we consider the weakly nonuniform fluctuating field $\mathbf{E}'$, $a,a'\ll 1$, the effect of the RRF action on the atoms can be strong, particularly, due to the ability of RRF to retain the sign on macroscopic spatial scales $L\gg \lambda$. In the next section we will see in more detail how the rectified gradient force is calculated and how the RRFs act on the tripod-type atoms.

\section{Fokker-Planck equation (FPE)\label{}}
One obtains first of all the kinetic equation FPE, describing the DF $f$ evolution at times
\begin{equation}\label{f31}
t\agt\tau_r\sim\omega_r^{-1}=\left( \frac{\hbar k^2}{m} \right)^{-1}\gg R_1^{-1},\quad R_1=\frac{R}{\cos\beta}.
\end{equation}
\par

Use is made of the procedure of the adiabatic elimination of the internal degrees of freedom of the atom which is well-developed in the theory of the resonant light pressure \cite{Kazantsev1990, Minogin1987, Metcalf2002}, i.e. in our case, the elimination of the variables $q_i$, included in Eqs. (\ref{f11}), (\ref{f12}). In fact, it implies the separation of the fast processes of the redistribution of the tripod-type atom over the quantum states (occurring at times $t\sim R^{-1}$) from the slow processes associated with the translational motion (occurring at times $t\gg R^{-1}$).
\par

One should take into account the smallness of the parameter $a_1=2(a+a')/(1+a^2+a')$ (according to the assumption on the weak non-homogeneity of the field $\mathbf{E}'$, at $a,a'\ll 1$, see Eqs. (\ref{f21}), (\ref{f28})) and quasi-classicality parameter  $\varepsilon=(\hbar k/ms)\ll 1$. Moreover, we restrict ourselves to the case of the slow atoms:
\begin{equation}\label{f32}
s^2< v_c^2,
\end{equation}
where $v_c\sim R_1/k$ has the meaning of a so-called capture velocity, i.e. of the characteristic velocity, which determines the area of the most efficient RRFs action in the problem under consideration: $|\mathbf{v}|\ll v_c$. It can be shown that outside this region the rectified gradient force (RGF) and coefficient of the light-induced friction fall rapidly proportional to $(v_c/v)^2$. The physical meaning of formula $v_c\sim R_1/k=R/k\cos\beta$ is very simple: $v_c$ is a typical velocity at which the atom travels a distance of the order of microscopic spatial scale $\lambda_M=\lambda/2\cos\beta$ during the characteristic time  $\sim R^{-1}$, i.e. $v_c\sim\lambda_M R$; the slow atoms travels a small distance (compared with the $\lambda_M$) during this time. That is why $v_c$ value can be adjusted by geometric parameter changing: $\cos\beta$ (this effect is described in the paper [20], Section 2.3 for two-level atoms and bichromatic field). It is worth noticing that in the case of the rectification schemes considered in \cite{Kazantsev1989, Krasnov1994, Krasnov2004} $v_c$ is determined only by the spontaneous relaxation rate and does not depend on the field intensity: $v_c\sim\gamma/k$. The quasi-stationary solution of Eqs. (\ref{f11}), determining the relation between $q_i$ and $f$ at $t\gg\gamma^{-1},R^{-1}$ is presented as an expansion over the small parameters of the problem ($\varepsilon$, $a_1$, ($v/v_c$)):
\begin{equation}\label{f33}
q_i=\hat{q}_i^{(0)}f+\hat{q}_i^{(1)}f+...,
\end{equation}
where $\hat{q}_i^{(n)}$ are linear operators.
\par

Using linear approximation over the velocity $\mathbf{v}$ and parameters $\varepsilon$, $a_1$ (i.e. taking into account in expansion (\ref{f33})only the dominant terms, proportional to the small parameters to the power not exceeding the first power) the following expression for $q_i$ was obtained:
\begin{equation}\label{f35}
q_i\simeq\frac{f}{4\chi+3}+\sum_{\alpha=0}^3\hat{q}_{i\alpha}f,
\end{equation}
\begin{equation*}
\hat{q}_{i0}=-\frac{a_1 f}{4\chi+3} [ 2(2\chi+1)P_i(\mathbf{r})-P_l(\mathbf{r})-P_j(\mathbf{r}) ],
\end{equation*}
\begin{eqnarray*}
\hat{q}_{i1} & = &\frac{a_1 f}{(4\chi+3)^2R} [ 2(6\chi^2+8\chi+3)\dot{P}_i(\mathbf{r}) \\
&& - (4\chi^2+8\chi+3)(\dot{P}_l(\mathbf{r})+\dot{P}_j(\mathbf{r})) ],
\end{eqnarray*}
\begin{equation*}
\hat{q}_{i2}=\mathbf{N}\cdot\frac{\partial}{m\partial\mathbf{v}}\frac{1}{(4\chi+3)^2R}, \quad
\hat{q}_{i3}=-\frac{1}{\gamma(4\chi+3)}\frac{d}{dt},
\end{equation*}
\begin{eqnarray*}
\mathbf{N} & = & (4\chi^2+5\chi+2)\frac{\hbar\nabla|V_{i1}(\mathbf{r})|^2}{\Delta_i} \\
    &&- (2\chi+1)\nabla
    \left( \frac{\hbar|V_{l1}(\mathbf{r})|^2}{\Delta_l}+\frac{\hbar|V_{j1}(\mathbf{r})|^2}{\Delta_j} \right),
\end{eqnarray*}
where $l\ne j\ne i$, $\dot{P}_i(\mathbf{r})=\mathbf{v}\cdot\nabla P_i(\mathbf{r})$, $\chi=R/\gamma$. The second term in Eq. (\ref{f35}) describes the effect of the non-homogeneous distribution of the atoms over the quantum states due to the spatial modulation (with the period $\sim \lambda$) of the transition rates  $R_i(\mathbf{r})$ and light shifts $|V_{i1}(\mathbf{r})|/\Delta_i$. It is important that non-local effects (the effects on non-adiabaticity in the atom response to the action of external fields) are taken into account, since the terms $\hat{q}_{i\alpha}f$ ($\alpha=1-3$) in Eq. (\ref{f35}) are determined by the field intensity gradients and are connected with the atom motion and recoil effect.
\par

Substituting expression (\ref{f35}) for  $q_i$ in Eq. (\ref{f12}), retaining terms up to the second order in the parameter $\varepsilon=\hbar k/ms$ and averaging the equation for DF over the small-scale spatial oscillations with a period of the order of the light wave-length $\sim \lambda$, one obtains after some transformations the following FPE for the Wigner distribution function averaged over the microoscillations, $\bar{f}$:
\begin{equation}\label{f36}
\frac{d\bar{f}}{dt}+\frac{\partial}{m\partial\mathbf{v}}(\mathbf{F}_R+\mathbf{F}_{R1})\bar{f}= D\left(\frac{\partial}{\partial\mathbf{v}}\right)^2\bar{f} + \sum_i D_{Ri}\frac{\partial^2 \bar{f}}{\partial v_i^2},
\end{equation}
where $\mathbf{F}_R$ does not depend on the atom velocity and has the meaning of the rectified gradient force (RGF),
\begin{equation*}
\mathbf{F}_R=-\sum_{i=x,y,z}\left< \frac{\hbar\nabla|V_{i1}(\mathbf{r})|^2}{\Delta_i}\hat{q}_{i0}(\mathbf{r}) \right>_s,
\end{equation*}
$\mathbf{F}_{R1}$ is linear-in-velocity RRF (the retarded gradient force by the terminology of [1], which appearance is associated with hysteresis in the response of the moving atom on external field):
\begin{equation*}
\mathbf{F}_{R1}=-\sum_{i=x,y,z}\left< \frac{\hbar\nabla|V_{i1}(\mathbf{r})|^2}{\Delta_i}\hat{q}_{i1}(\mathbf{r},\mathbf{v}) \right>_s,
\end{equation*}
$D$, $D_{Ri}$ are the diffusion coefficients (determined by the terms $\hat{q}_{i2}f$ and $\hat{q}_{i3}f$ in the expansion (33) and also by terms of the right side of the Eq.\,(12)). Note, that in our approximation (\ref{f35}) we neglect small additives to RRFs, which have the order of magnitude $a_1^3$ and $(v/v_c)^2a_1$. The coefficients of Eq. (\ref{f36}) do not contain components oscillating (in space) on microscopic spatial scales and can change only on macroscopic scales $\agt L$. This is explained by the fact that relative differences of population $\hat{q}_i=q_i/f$ (See Eqs.\,(27),\,(33)) contain oscillating components with spatial frequencies close to spatial frequencies of gradient force oscillations $\propto\nabla|V_{i1}(\mathbf{r})|^2$. The condition of the correctness of the implemented procedure of averaging over the small-scale spatial oscillation (which is quite similar to the case of the two-level atoms \cite{Krasnov1994}) is the following:
\begin{equation}\label{f37}
\frac{ms^2}{2}\sim T\gg\frac{\hbar|V_{1i}(\mathbf{r})|^2}{\Delta_i(4\chi+3)},
\end{equation}
and implies that the effective temperature $T$ (in energy units) of the atoms is considerably higher than the depth of the macroscopic potential wells  produced by the oscillating (with a period of $\sim\lambda$) gradient force with the potential
\begin{equation*}
U_g(\mathbf{r})=\sum_{i=x,y,z}\frac{\hbar|V_{1i}(\mathbf{r})|^2}{\Delta_i(4\chi+3)}.
\end{equation*}
\par

In the situations considered here the condition is always satisfied and the  rapidly oscillating component $\tilde{f}$ of DF is a small correction to $\bar{f}$: $(\tilde{f}/\bar{f}\sim a_1\ll 1)$.
\par

Given all the detunings $\Delta_i>0$, the parameters $\xi_i<0$ ($i=x,y,z$) and the Stark shifts of the energy levels induced by the coherent fields with mutually orthogonal polarization directions are equal, i.e the the following relations hold:
\begin{equation}\label{f38}
\frac{V_i^2}{\Delta_i}=\gamma\sqrt{\frac{I}{I_s}g},\quad g=\left( \frac{V_x}{\Delta_x} \right)^2, \quad \frac{I}{I_s}=\frac{V_x^2}{\gamma^2},
\end{equation}
where $I=I_x$ is the intensity of the light waves forming the field component $E_{x1}\mathbf{e}_x$, $I_s=\hbar\omega_0k^2\gamma/6\pi$ is the intensity of the optical radiation saturating the quantum transition. In this case the explicit expressions for the forces included in Eq. (\ref{f36}), have the following compact form:
\begin{equation}\label{f39}
\mathbf{F}_R=-\nabla U_R+\mathbf{F}_{R0}, \quad \mathbf{F}_{R1}=-m\sum_{i=x,y,z}\varkappa(r_i)v_i\mathbf{e}_i,
\end{equation}
\begin{equation*}
U_R=U_0\sum_{i=x,y,z}\cos\left( \frac{2\pi}{L_1}r_i-\eta_1 \right),
\end{equation*}
\begin{equation}\label{f40}
U_0=\hbar\gamma\frac{kL_1}{2\pi}\frac{a_1G}{(1+b)}A(\chi),\quad G=\sqrt{\frac{I}{I_s}g}\cos\beta,
\end{equation}
\begin{equation}\label{f41}
\mathbf{F}_{R0}=-\hbar k\gamma\frac{a_1bG}{(1+b)}\sin(\eta_1-\zeta_1)A(\chi)\sum_{i=x,y,z}\mathbf{e}_i,
\end{equation}
\begin{eqnarray}\label{f42}
    \varkappa(r_i)&=&\frac{2(6\chi^2+8\chi+3)}{(4\chi+3)^3\chi}\frac{a_1G\cos\beta}{(1+b)}\omega_R \nonumber \\
    &&\times \left[ b\cos(\eta_1-\zeta_1)+\cos\left( \frac{2\pi}{L_1}r_i-\eta_1 \right) \right] ,
\end{eqnarray}
where $r_i=(\mathbf{r}\cdot\mathbf{e}_i)$, $v_i=(\mathbf{v}\cdot\mathbf{e}_i)$, $L_1=L/\cos\beta$, $A(\chi)=(2\chi+1/(4\chi+3)^2$.
\par

One can see from Eqs. (\ref{f39})--(\ref{f41}) that RGF $\mathbf{F}_R$ contains in the general case $(\eta_1\ne\zeta_1)$ the component $\mathbf{F}_{R0}$ which does not depend on the spatial coordinates $\mathbf{r}$. This is the manifestation of the total rectification effect appearing due to a special choice of the wave vectors (see Eq.(\ref{f25}) and Fig. 1). Another component of RGF, $-\nabla U_R$, gives rise to an optical superlattice, a system of periodically (with a period $L_1\gg\lambda$) distributed potential wells for the atoms. Though the force $\mathbf{F}_{R0}$ is likely to have special practical applications; here, one is interested in a certain problem of the \emph{3D} localization and cooling of the atoms, and, therefore, should restrict oneself to a special case of choosing the phases of the fields $\eta_1$ and $\zeta_1$ and the parameter $b=a'/a$ in Eq. (\ref{f42}):
\begin{equation}\label{f43}
\eta_1=\zeta_1=\pi,\quad b>1.
\end{equation}
Thus, $\mathbf{F}_{R0}=0$, and  $\mathbf{F}_{R1}$ is a linear-in-velocity friction force, determined  by the spatially nonuniform strictly positive friction coefficients
\begin{equation}\label{f44}
\varkappa_i=\varkappa(r_i)>0.
\end{equation}
\par

Interestingly, in this case expressions (\ref{f39})--(\ref{f42}) coincide with the expressions for RRFs, obtained for the tripod-type ion by another method in  \cite{Krasnov2009}.
\par

Thus, under the conditions considered RRFs induce the \emph{3D} dissipative cubic superlattice, capable of simultaneously cooling and trapping atoms. However, the kinetics of the atom cooling and trapping considerably depends on the values of the velocity diffusion coefficients $D$ and $D_{Rj}$ (determining the process of the atom heating competing with the atom cooling). Here, the following explicit expressions were obtained:
\begin{equation*}
D=D_{s1}+D_1,
\end{equation*}
\begin{equation*}
D_{s1}\simeq\left( \frac{\hbar k}{m} \right)^2\frac{\gamma\chi}{(4\chi+3)2}, \quad D_1\simeq\left( \frac{\hbar k}{m} \right)^2\gamma\frac{\chi(4\chi+2)}{(4\chi+3)4},
\end{equation*}
\begin{equation}\label{f45}
D_{Rj}=\left( \frac{\hbar k}{m} \right)^2\gamma \frac{8\chi^3+16\chi^2+11\chi+3}{(4\chi+3)^3\chi}G^2=D_R,
\end{equation}
where $j=x,y,z$, $D_{s1}$ is the diffusion coefficient conditioned by the recoil at spontaneous transitions, $D_1$ is the coefficient of the induced diffusion connected with the action (on the atoms) of the fluctuating field $\mathbf{E}'$\footnote{Though under the conditions considered, the light pressure force conditioned by the field $\mathbf{E}'$ is equal to zero (according to Eq. (\ref{f30})),its fluctuations are not equal to zero.} ( its order of magnitude coincides with that of the diffusion coefficient in the case of the two-level atoms in the field of the standing wave with the fluctuating phase \cite{Kazantsev1978}), $D_R$ is the coefficient of the induced diffusion connected with the quantum fluctuations of the gradient force. When writing Eqs.\,(43) we neglect small corrections of the diffusion coefficient $\sim a_1^2$, $(s/v_c)a_1^2$.
\par

The distinct feature of the considered \emph{3D} scheme of the mechanical action of the light on the tripod-type particles is the simultaneous presence of several independent governing parameters: $\chi$, $G$, $L_1$, $\cos\beta$, $a_1$, $b$, $\zeta_1$, $\eta_1$. The specific choice of these parameters determines the values of the force and diffusion coefficients of FPE, and, consequently, various regimes of the atom localization and cooling. Now we shall describe some characteristic situations, considering  Eq. (\ref{f43}) to be true and using Eqs. (\ref{f39}), (\ref{f40}), (\ref{f42}) and (\ref{f45}).
\par

\subsection{Strong fields: $G\gg1$, $\chi\gg1$ \label{}}

In this case RGF $F_R\sim F_sGa_1/\chi$ and can considerably exceed the maximum value of the so-called spontaneous light pressure force $F_s=\hbar k\gamma/2$ \cite{Kazantsev1990}, if $G\gg \chi/a_1$. The capture velocity $v_c$ can exceed the characteristic value $v_{c0}=\gamma /k$ by a large factor $(\chi/\cos\beta)\gg 1$. The friction coefficient can exceed the characteristic value $\omega_R$: $\varkappa\sim\omega_R a_1\cos\beta G/\chi^2\gg\omega_R$, if $G\gg\chi^2/a_1\cos\beta$. The depth of the macroscopic potential wells $\Delta U=2U_0$ considerably exceeds the depth of the microscopic potential wells $U_g$: $\Delta U\sim U_g(Ga_1L/\lambda\chi)\gg U_g$, in case $(a_1L/\lambda)\gg\chi/G$.
\par

Thus, high values of RGF and capture velocity $v_c$ can be achieved as well as super-fast cooling rates, if $G\gg\chi^2$. However, in this regime the atom heating rate is also high. Really, comparing the velocity diffusion coefficient $\bar{D}=D+D_R$ (determining the heating rate) with the characteristic value of the diffusion coefficient in a traveling plane wave \cite{Kazantsev1990, Minogin1987, Metcalf2002}, $D_0=(\hbar k/m)^2\gamma$, one obtains:
\begin{equation*}
\bar{D}=D_0\left( \frac{G^2}{8\chi}+\chi \right) \gg D_0.
\end{equation*}

\subsection{Strong coherent field and not very strong partially coherent field: $G\gg1$, $\chi\alt1$ \label{}}

One has the following relations: $F_R\sim F_s Ga_1\gg F_s$ (if $Ga_1\gg 1$), $\Delta U\sim U_g(La_1/\lambda)\gg U_g$ (if $L\gg\lambda/a_1$), $\varkappa\sim\omega_RGa_1\cos\beta/\chi\gg\omega_R$ (if $G\gg\chi/a_1\cos\beta$), $v_c=v_{c0}$ (if $\chi\sim1$ и $\cos\beta\approx1$) and $v_c\gg v_{c0}$ (if $\chi\sim 1$ and $\cos\beta\ll1$), $\bar{D}\sim D_0G^2/\chi\gg D_0$.
\par

This implies that in this case it is also possible to reach super-high values of   RGF, high values of the capture velocity $v_c$ (but only by means of the proper choice of the geometric factor: $\cos\beta\ll1$), super-high cooling rates and intensive diffusion in the velocity space. The relatively small intensities of the fluctuating field $\mathbf{E}'$ make it significantly different from the case A.
\par

A similar regime of the mechanical action of the coherent bichromatic light on the two-level atoms is described, for example, in: \cite{Kazantsev1989, Krasnov1994}, however, these descriptions cannot be applied directly for the case  of the \emph{3D} configuration (considered here) of the optical fields and tripod-type atoms.
\par

\subsection{Weak fields: $G\sim\chi\ll1$ \label{}}

There exist the following estimates: $F_R\sim F_s a_1G$, $\Delta U\sim U_g a_1L/\lambda$, $v_c=v_{co}\chi/\cos\beta$, $\varkappa\sim\omega_Ra_1\cos\beta$, $\bar{D}\sim D_0G$.
\par

The most significant peculiarity of this case is that the friction coefficient does not change with the decreasing intensity of the fields, but the coefficient of the velocity diffusion decreases. In other words, it appears possible to considerably suppress the radiation force fluctuations when maintaining the high cooling rate. A similar physical situation appears in the famous \emph{1D} model of the polarization gradient cooling (see \cite{Cohen1990} and references) and it is a prerequisite of achieving the Sub-Doppler temperatures.

\section{Cooling, spatial diffusion and localization of atoms. \label{}}

An important feature of the Brownian motion of atoms, described by FPE (\ref{f36}) for $\varkappa >0$ is the overdamped character of this motion:
\begin{equation}\label{f46}
\frac{s}{\varkappa}\sim\lambda_r\ll L_1,
\end{equation}
where $\lambda_r$ has the meaning of the effective free-path length of the atoms in viscous "fluids" of photons. A similar si\-tuation occurs in OM \cite{Chu1985}, if $L_1$ denotes its dimensions. In the problem considered $\varkappa\agt\omega_Ra_1$ (see section IV), therefore, assuming $k\simeq 10^5$ $cm^{-1}$, $\gamma\sim10^{-8}$ $s^{-1}$, $m\sim100$ $amu$, $s\sim\sqrt{\hbar\gamma/m}$, $a_1=0.2$, one has the following: $\lambda_r\alt10^{-3}$ $cm$. Thus, condition (\ref{f46}) of the strong dissipativity of the optical superlattice is well satisfied, if the period is $L_1\agt0.01$ $cm$.
\par

Assuming inequality (\ref{f46}) to be satisfied let us consider the evolution of the atom ensemble at times  $t>\tau_d\sim\varkappa^{-1}(L_1/\lambda_r)^2\gg\tau_0\sim\varkappa^{-1}$, where (according to the general properties of FPE \cite{Risken1989}) $\tau_0$ is the time necessary for the local quasi-stationary velocity distribution to be achieved, $\tau_d$ is the characteristic time of developing a much slower (diffusion) process of changing the atom density, $n(\mathbf{r},t)$, in the configuration space ($\mathbf{r}$).
\par

Then, it appears possible to present the solution of FPE (\ref{f36}) in the following form (using instead of the variables $(\mathbf{r},\mathbf{v})$ the variables $(\mathbf{r},\mathbf{c}=\mathbf{v}-u(\mathbf{r},t))$, where $\mathbf{u}(\mathbf{r},t)$ is the macroscopic (directed) velocity of the atom motion):
\begin{equation}\label{f47}
f=nf_0+f_1,
\end{equation}
\begin{equation}\label{f47a}
\langle f_1\rangle_c,\quad \langle f_1\mathbf{c}\rangle=0, \quad f_1\ll nf_0,
\end{equation}
where the angular brackets  $\langle\cdots\rangle_c$ denote the integration over the velocities of the chaotic motion $\mathbf{c}$,
\begin{equation}\label{f48}
f_0=\prod_{i=x,y,z}f_i,\quad f_i=\left( \frac{m}{2\pi T_i} \right)^{\frac{1}{2}}\exp\left[ -\frac{mc_i^2}{2T_i} \right],
\end{equation}
$T_i=T(r_i)=m(D+D_R)/\varkappa(r_i)=T_0\hat{T}_i$  has the meaning of the effective local atom temperature (in energy units), characterizing the chaotic atom motion along the axis $i$ $(i=x,y,z)$, $T_0$ is the space averaged effective atom temperature, $\hat{T}_i=\hat{T}(r_i)$ is the function describing the effective temperature dependence $T_i$ on the corresponding spatial coordinate $r_i$. Explicit expressions for them are obtained from Eqs. (\ref{f42}), (\ref{f45}), taking into account Eq. (\ref{f43}):
\begin{equation*}
\hat{T}(r_i)=\frac{b_1}{b-\cos\left( \frac{2\pi r_i}{L_1} \right)},\quad b_1=\sqrt{b^2-1},
\end{equation*}
\begin{equation}\label{f49}
T_0=\frac{m(D+D_R)}{\varkappa_0}, \; \varkappa_0=\frac{2(6\chi^2+8\chi+3)a_1b_1G\cos\beta}{(4\chi+3)^3\chi(1+b)}\omega_R.
\end{equation}
\par

In the considered approximation, $\lambda_r\ll L_1$ the density $n(\mathbf{r},t)$ and macroscopic velocity $\mathbf{u}(\mathbf{r},t)$ are governed by the equations (continuity equation and the force balance equation):
\begin{equation}\label{f50}
\frac{\partial n}{\partial t}+{\rm div} (n\mathbf{u})=0,
\end{equation}
\smallskip
\begin{equation}\label{f51}
\frac{\partial(nT_i)}{\partial r_i}+\frac{\partial U_R}{\partial r_i}+m\varkappa_i u_i=0,
\end{equation}
where $\varkappa_i=\varkappa(r_i)$, $u_i=(\mathbf{e}_i\cdot\mathbf{u})$. It can easily be proved that $f_0$ satisfies the equation ($c_i=(e_i\cdot\mathbf{c})$):
\begin{equation*}
\hat{L}f_0=0,\quad \hat{L}=\sum_i \frac{\partial}{\partial c_i}\left[ \varkappa_i \left( c_i+\frac{T_i}{m}\frac{\partial}{\partial c_i} \right) \right],
\end{equation*}
and a small correction $f_1$ to $nf_0$ in the first approximation satisfies the equation
\begin{widetext}
\begin{equation}\label{f52}
\hat{L}f_1=f_0n
\left(
    \sum_{i\ne j}\frac{mc_ic_j}{T_j}\frac{\partial u_j}{\partial r_i} +
    \sum_i\left( \frac{mc_i^2}{T_i}-1 \right) \left( \frac{\partial u_i}{\partial r_i} + u_i\frac{\partial T_i}{2T_i\partial r_i} \right) +
    \sum_i \frac{\partial T_i}{2T_i\partial r_i}c_i\left( \frac{mc_i^2}{T_i}-3 \right)
\right).
\end{equation}
\end{widetext}
Expanding the solution of Eq. (\ref{f52}) into the eigenfunctions of the Fokker-Plank operator $\hat{L}$ and taking into account  Eq. (\ref{f50}), (\ref{f51}) one can obtain a quite strict estimate
\begin{equation*}
\frac{f_1}{nf_0}\sim\frac{\partial u_j}{\partial r_i\varkappa_0}, \quad \frac{\partial T_i}{T_i\partial r_i\varkappa_0} u_i\sim \left( \frac{\lambda_r}{L_1} \right)^2 \ll 1.
\end{equation*}
\par

Thus, it is seen from Eq. (\ref{f48}), (\ref{f49}), that at $t\gg\varkappa^{-1}$ and condition (\ref{f46}) the atom distribution over the velocities (of the chaotic motion) is, generally speaking, spatially nonuniform and anisotropic (with the exception of the points $\mathbf{r}'$ located on the bisectors of the unit cells of the optical superlattice, in which $\cos(2\pi r_i'/L_1)=\cos(2\pi r_j'/L_1)$). However, within the limit $b\gg 1$, $\hat{T}\approx1$ and $T_i\simeq T_0$, DF $f_0$ is the equilibrium Maxwell distribution with the temperature $T_0$.
\par

Further, combining Eqs. (\ref{f50})--(\ref{f51}) one obtains the Smoluchowski equation (SE), describing the atom diffusion in the field RGF
\begin{equation}\label{f53}
\frac{\partial n}{\partial t}=\sum_{i=x,y,z}\frac{\partial}{\partial r_i}\frac{1}{m\varkappa_i} \left[ T_i\frac{\partial n}{\partial r_i} + n\frac{\partial}{\partial r_i}(U_i+T_i) \right],
\end{equation}
where $U_i=-U_0\cos(2\pi r_i/L_1)$.
\par

The difference of SE (\ref{f53}) from the classical SE (well-known in the theory of the Brownian motion \cite{Chandrasechar1943}) – is the spatial non-uniformity of the friction coefficients $\varkappa_i=\varkappa(r_i)$ and effective temperatures $T_i=T(r_i)$.
\par

Taking into account the form of the obtained SE (\ref{f53}) a very significant and interesting circumstance appears: in the model considered the macroscopic (diffusion) particle motions along the axes of the Cartesian coordinate system can entirely be separated (i.e. they can be independent from each other). This is revealed in the stationary solution of SE (\ref{f53}) being the product of the quasi-Boltzmann distributions:
\begin{equation}\label{f54}
n=\prod_{i=x,y,z}\frac{\psi_i}{\hat{T}(r_i)}\exp \left[ - \frac{U_{eff}(r_i)}{T_0} \right],
\end{equation}
where  $\psi_i$ - are the constants, and the effective potential $U_{eff}$ is determined by the expression
\begin{equation}\label{f55}
U_{eff}(r)=-\frac{U_0}{b_1}\left( b\cos\left[ \frac{2\pi r}{L_1} \right] -\frac{1}{2}\cos^2\left[ \frac{2\pi r}{L_1} \right] \right),
\end{equation}
and describes the potential wells with the depth $\Delta U=2U_0b/b_1$.
\par

The condition of the deep particle localization has the form:
\begin{equation}\label{f56}
\Gamma=\exp\left[ \frac{\Delta U}{T_0} \right]\gg1.
\end{equation}
\par

A prerequisite of satisfying condition (\ref{f56}) is the presence of the large
multiplier $L_1/\lambda\agt 10^4$ (at $L_1\sim 1$ $cm$) in the expression for $U_0$ (\ref{f40}). However, stationary localized solutions of SE (\ref{f53}) are strong idealizations due to the irreversible diffusion particle escape from the interception area of the real laser beams with the finite transverse dimensions. Consider a more realistic model allowing one to take this effect into account and to show the existence of the spatially localized quasi-stationary (long-lived) states of the atom ensemble. Considering, for cirtainty, a single cubic cell of the superlattice (described by equation (\ref{f40}) at $\eta_1=\zeta_1=\pi$) with the centre in the point $\mathbf{r}=0$ as the dissipative optical trap (DOT) for the atoms, we assume that the absorbing boundary condition is satisfied: $n_\Sigma=0$ \cite{Risken1989} on the boundaries $\Sigma$ of this cell. It means that the particles are removed from the DOT as soon as they reach its boundaries.
\par

Then, the long-lived localized states of the atom ensemble (corresponding to the lowest (the slowest) diffusion mode), can still be presented as (\ref{f54}), assuming that the multipliers $\Psi_i=\bar{\Psi}(r_i)\exp[-t/\tau_1]$, where $\bar{\Psi}(r)$ is the eigenfunction, corresponding to the lowest eigenvalue $\lambda_1=1/\tau_1$ of the Sturm-Liouville boundary problem (SLP)
\begin{equation*}
\left( \hat{H} + \frac{1}{\tau_1}\frac{\Phi}{\hat{T}(r)} \right)\bar{\Psi}(r)=0, \quad \bar{\Psi}\left(\pm \frac{L_1}{2} \right)=0,
\end{equation*}
\begin{equation}\label{f57}
\hat{H}=\frac{d}{dr}\left( \frac{T_0}{m\varkappa}\Phi\frac{d}{dr} \right),\quad \Phi=\exp\left[ -\frac{U_{eff}(r)}{T_0} \right].
\end{equation}
\par

Using (with slight modifications) the results of the analysis of a similar SLP (in the case of $\Gamma\gg1$), made in \cite{Krasnov2009}, one obtains the following expression for the lifetime $\tau=\tau_1/3$ of the atoms in DOT
\begin{equation}\label{f58}
\tau\simeq\frac{\tau_d}{3\pi\hat{T}(0)\ln\Gamma}\exp\left[ \frac{\Delta U}{T_0} \right],
\end{equation}
where $\tau_d\simeq(b+1)L_1^2m\varkappa_0/T_0b_1$ is the characteristic diffusion time in DOT in the absence of RGF. Thus, condition (\ref{f56}) is simultaneously the condition of the long-term atom confinement in DOT (for a time considerably longer than the diffusion time $\tau_d$). Note here that at $\ln\Gamma\gg1$, $\bar{\Psi}(r)\approx{\rm const}$ almost everywhere except for the narrow regions (with the width of $\sim L_1/\sqrt{\ln\Gamma}$) near the boundaries $\Sigma$ of the trap.
\par

Consider a number of characteristic properties of the spatially localized states of the atom ensemble resulting from the constructed model.
\par

A limitation for the effective mean temperature $T_0$ of the localized particles results from Eqs. (\ref{f45}), (\ref{f49})
\begin{equation}\label{f59}
T_0>\frac{\hbar\gamma}{2a_1 b_1}\frac{b+1}{\cos\beta}G.
\end{equation}
\par

This inequality, on the one hand, pre-determines satisfying important condition (\ref{f37}) due to the smallness of the parameter $a_1$: $(2a_1/3)\ll1$. On the other hand, from inequality (\ref{f59}) it follows that reaching Sub-Doppler temperatures $T_0\ll T_D\sim\hbar\gamma$ is possible only at rather small values of the coherent field intensities (i.e. at $G<(2a_1b_1/b+1)\ll1$). It seems quite natural since in the limit $G\ll1$ and $\chi\ll1$:
\begin{equation}\label{f60}
T_0=BT_D\left( G+\frac{3\chi^2}{G} \right)\geqslant 2\sqrt{3}B\chi T_D=T_m,
\end{equation}
where $T_D=\hbar\gamma$ is the characteristic temperature, determining the Doppler limit of the atom cooling \cite{Kazantsev1990, Minogin1987, Metcalf2002}, $B=(b+1)/2a_1b_1\cos\beta$ and the equality is achieved at $G=\sqrt{3}\chi$. Thus, the Sub-Doppler cooling regime can be implemented at $\chi<a_1b_1\cos\beta/(b+1)\sqrt{3}$. On the other hand, the parameter $\chi$ (at $G=\sqrt{3}\chi$) should be limited by lower bound: $\chi >\omega_R\sqrt{3}b_1/\gamma a_1(b+1)$, to sa\-tisfy condition (\ref{f32}), $s^2<v_c^2=R^2/k^2\cos^2\beta$. Regarding this circumstance and Eq. (\ref{f60}) one obtains the estimate of the lower limit $T_m'$ of the effective mean temperature of the described localized states of the atom ensemble
\begin{equation}\label{f61}
\frac{T_m'}{T_D}\sim\frac{3\omega_R}{\gamma a_1^2}.
\end{equation}
It follows from Eq. (\ref{f61}) that $T_m'/T_D\ll1$, if $(\omega_R/\gamma)\ll a_1^2$. Note, that the typical values of $\omega_R/\gamma$ are $\alt 10^{-3}$. Reaching Sub-Doppler temperatures easily corresponds to the condition of the long-term deep spatial atom localization (\ref{f56}) making the proper choice of the period $L_1$ of the optical super-lattice (induced by RGF):
\begin{equation}\label{f62}
\ln\Gamma=\frac{4a_1^2b\cos\beta}{9(b+1)^2}\cdot\frac{L_1}{\lambda}>1.
\end{equation}
\par

Now let $G\gg 1$ and $\chi\alt 0.1$. This case corresponds to Super-Doppler temperatures of the localized atoms:
\begin{equation}\label{f63}
T_0=T_DBG\gg T_D.
\end{equation}
Moreover, the condition of the long-term deep localization is also expressed by inequality  (\ref{f62}). It is necessary to take into account that the values of the parameter $G$ are limited by condition (\ref{f32}): $G<\chi^2\gamma/\omega_RB\cos^2\beta$. Then, from Eq. (\ref{f63}) one can obtain the estimate (at $\chi\sim 0.1$) of the maximum temperatures $\max T_0$:
\begin{equation}\label{f64}
\frac{\max T_0}{T_D}\simeq\frac{0.01\gamma}{\omega_R\cos^2\beta},
\end{equation}
which can be reached in this regime of the optical field influence on the  tripod-type atoms.
\par

By smooth and slow variation of the saturation parameters $\chi$ and $G$ (proportional to the intensities of the fluctuating and coherent field components) it is possible to implement a continuous transition from ``Super-Doppler spatially localized states'' of the atom ensemble (with the temperatures $T_0\sim T_{sup}\gg T_D$) to ``Sub-Doppler states'' (with the temperatures $T_0\sim T_{sub}\ll T_D$)) or to any  ``intermediate'' state with the necessary temperature value from the interval $[T_{sub}T_{sup}]$: $T_{sub}<T_0<T_{sup}$.
\par

However, the changes of the parameters $G$ and $\chi$ are to be matched (for example, by means of setting a certain relation between them, $G=G(\chi)$), in order   not to violate the conditions of the stable spatial atom localization (\ref{f56}) and (\ref{f32}), at any values of the parameters mentioned. Let us give an example demonstrating the possibility of the continuous monotonic temperature change of the trapped atoms from $T_{sup}$ up to $T_{sub}$ without violating the conditions of the stable localization.
\par

Let $T_D=10^{-3}$ $K$ ($\gamma\approx1.31\cdot10^8$ $s^{-1}$), the wave number $k\approx 10^5$ $cm^{-1}$, $m\approx 200$ $amu$, $L_1=1.2$ $cm$, $a_1^2\approx 0.1$, $\cos\beta=0.1$, $b=1.5$, and the matching law is
\begin{equation}\label{f65}
G=\sqrt{3}\delta\left(\exp\left[ \frac{\chi}{\delta} \right] -1 \right),
\end{equation}
where $\delta=0.015$. Thus, at $\chi\gg\delta$ the parameter $G>1$, and at $\chi\ll\delta$, the parameter $G\approx\sqrt{3}\chi$ (which corresponds to the optimal relation between $G$ and $\chi$ for reaching Sub-Doppler temperatures).
\par

\begin{figure}
\includegraphics{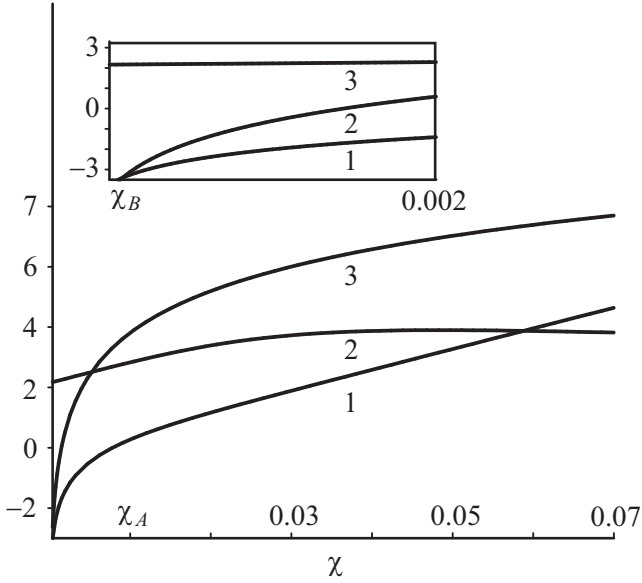}
\caption{
The dependence of the characteristics of the localized states of the atom ensemble on the governing parameter $\chi$ at continuous transition from ``Super-Doppler states'' ($T_0\gg T_D$) to ``Sub-Doppler states'' ($T_0\ll T_D$) and according to the matching law (\ref{f65}): curve 1 --- $\ln (T_0(\chi)/T_D)$; curve 2 --- $0.2\times\ln\Gamma(\chi)$; curve 3 --- $\ln(mv_c^2(\chi)/T_D)$; $T_D$ is the Doppler cooling limit, $\gamma\approx1.31\cdot10^8$ $s^{-1}$, $k\approx 10^5$ $cm^{-1}$, $m=200$ $amu$, $L_1=1.2$ $cm$, $a_1^2\approx0.1$, $\cos\beta=0.1$, $b=1.5$, $T_0(\chi_A)=T_D$, $\chi_A\approx 0.01$. Inset --- the details of the same curves in the Sub-Doppler temperature region $(\chi\ll\chi_A)$; $T_0(\chi_B)=mv_c^2(\chi_B)$, $\chi_B\sim 10^{-4}$.
\label{}}
\end{figure}

Given in Fig. 2 are the  dependencies (in a logarithmic scale) of the three main characteristics for the localized states of the atoms on the governing parameter $\chi$: the temperature, $\ln(T_0/T_D)$, capture velocity, $\ln(mv_c^2/T_D)$ and  localization parameter, $\ln\Gamma$ (See Eq.\,(48), comments to inequality (31) and Eq.\,(55)). One can see that when decreasing the governing parameter $\chi$ from the value $\chi\approx0.07$ ($G\approx 2.57$) to the value $\chi\sim0.001$ the temperature is monotonely decreased from the value $T_{sup}\approx100T_D\approx0.1$ $K$ to the value $T_{sub}\sim0.1T_D\approx100$ $\mu K$ (i.e. is decreased approximately by $1000$ times, ``running over'' all the intermediate values between $T_{sup}$ and $T_{sub}$). The conditions of stable localizations (\ref{f56}) and (\ref{f32}) are well satisfied. This, in particular, is revealed in the decay time of the localized states $\tau$ (see Eq. (\ref{f58})) in the given example exceeding the characteristic diffusion time $\tau_d\sim1$ $s$ not less than by two orders of magnitude for any value $\chi$ from the indicated range: $\tau>300\tau_d$. Attention should also be paid to a rather high temperature value $T_c\sim1$ $K$, corresponding to the value of the capture velocity $v_c$ on the right boundary of the $\chi$ change interval. This may have important practical significance for the effective solution of the problem of initial atom loading into the optical trap under consideration.

\section{Conclusion\label{}}

We have developed a \emph{3D} kinetic model, describing the mechanical action of non-monochromatic optical fields on tripod-type atoms. A distinctive feature of the proposed model is that light induced \emph{3D} atom kinetics is completely determined by the rectified radiation forces. Moreover, (under the conditions considered) the light induced atom motions in the three mutually orthogonal directions prove to be independent, which allows one to reduce the analysis of the \emph{3D} problem to the study of much more simple \emph{1D} problems and to obtain approximate analytical solutions of the kinetic equation for the Wigner distribution function of atoms $f(\mathbf{r},\mathbf{v},t)$.
\par

Another important peculiarity of the considered scheme of the mechanical action of light on the tripod-type atoms is the presence of several independent governing parameters. In particular, these are the parameters $G$ and $\chi$, proportional to the intensities of the coherent and partially coherent optical field components, phase shifts $\zeta_1$ and $\eta_1$, geometrical parameters (angular detunings $\delta_i$, $\delta_i'$ and the angle $\beta$ (see Fig. 1)) determining the optical field configuration. A purposeful choice of their values or relations between them makes it possible to set considerably different (desirable) regimes of the mechanical action of light on the atom motion.

 Note that, selection of the required controlling parameter combination is getting significantly easier due to obtaining of explicit analytical expressions in the proposed model (for the RRFs and the distribution function), which are convenient for analysis performance.
\par

Particularly, the developed theoretical model has been applied to analyze the problem of \emph{3D} optical cooling and trapping of the tripod-type atoms. We have made a description of the long-lived spatially localized atom ensembles (trapped into deep light-induced potential wells) and demonstrated the possibility of obtaining both ``Super-Doppler stable localized states'' (with the effective temperatures $T_{sup}$ considerably exceeding the Doppler cooling limit $T_D$ ($T_{sup}\gg T_D$) and high values of the capture velocity $v_c$ and  ``Sub-Doppler stable states'' (with the temperatures $T_{sub}$ considerably lower than the Doppler cooling limit ($T_{sub}\ll T_D)$). We have also shown the possibility of continuous transition between these states or reaching any other intermediate state with the effective temperature $T_{sub}<T_D<T_{sup}$ without violating the localization stability.
\par

Finally, it is worth noting that the tripod-type quantum transitions (which have already been used in the experiments on laser cooling and trapping) are characteristic, in particular, of the atoms $Na$ \cite{Nasyrov2001} (and the similar ones), as well as the ions $^{171}Yb^+$ and $^{199}Hg^+$ \cite{Tamm2000, Rafac2000}. Interestingly, for the transitions mentioned (in $Na$-like atoms) the usually accepted MOT model does not work; thus, in order to explain its work one should take into account the effects connected with the  existence of additional energy levels: magnetically induced level-mixing effects \cite{Nasyrov2001}. One can see that the considered scheme of all-optical cooling and trapping of the atoms does not require similar effects to be taken into account.
\par

Regarding the above-mentioned tripod-type ions attention is to be paid to the fact that recently there has been some interest in the development of all-optical methods of ion trapping \cite{Schneider2010}. Besides, such methods can be used for the solution of a new interesting problem, that of obtaining ultra-cold electron-ion plasma with resonant ions and its long-term confinement \cite{Krasnov2008, Krasnov2009}.
\par

Certainly, the application of the developed theoretical model to certain real atoms and ions may require additional analysis which would take into account, for example (if necessary) additional repumping laser fields.

\appendix*
\section{Explicit expressions for the terms of equations (\ref{f6}) and (\ref{f7}), describing the recoil effect}
Let us introduce an auxiliary function $Q_{ii}(\mathbf{r},\mathbf{v},t)$ and consider the relations between the function $\rho(\mathbf{r},\mathbf{v},t)$ and $f(\mathbf{r},\mathbf{v},t)$ и $q_{ii}(\mathbf{r},\mathbf{v},t)$ (where $i=x,y,z$):
\begin{equation}\label{a1}
Q_{ii}=f+q_{ii}-\sum_{l\ne i}q_{ll},
\end{equation}
\begin{equation}\label{a2}
\rho=\left(f-\sum_j q_{jj}\right)/4.
\end{equation}
\par

Then, the explicit expressions for the terms $\Lambda_i$, $B_{ij}$ in Eqs. (\ref{f6}), (\ref{f7}) can be written in the following form:
\begin{widetext}
\begin{equation}\label{a3}
    \Lambda_i=-\frac{\hbar i}{2m}
    \left[
        \frac{\partial\rho_i}{\partial\mathbf{v}}\cdot\sum_{l\ne i}\frac{\nabla|V_{e1}|^2}{\Delta_l}-
        \sum_{l\ne i}\frac{\partial q_{il}}{\partial\mathbf{v}}\cdot\nabla U_l-
        \frac{1}{2}\frac{\partial Q_{ii}}{\partial\mathbf{v}}\cdot\nabla U_i
    \right],
\end{equation}
\begin{equation}\label{a4}
    B_{ii}=\frac{\hbar i}{2m}
    \left[
        \sum_{l\ne i}
        \left(
            \frac{\partial\rho_l}{\partial\mathbf{v}}\cdot\nabla U_l^*+c.c.
        \right)
        +\frac{\partial Q_{ii}}{\partial\mathbf{v}}\cdot \frac{\nabla|V_{i1}|^2}{\Delta_i}
        +2\sum_{l\ne i}\frac{\partial\rho}{\partial\mathbf{v}}\cdot\frac{\nabla|V_{l1}|^2}{\Delta_l}
    \right],
\end{equation}
\begin{equation}\label{a5}
    B_{ij}=\frac{\hbar i}{2m}
    \left[
        \frac{\partial q_{ij}}{\partial\mathbf{v}}\cdot\nabla
            \left(
                \frac{|V_{i1}|^2}{\Delta_i}+\frac{|V_{j1}|^2}{\Delta_j}
            \right)
            -\left(
                \frac{\partial\rho_i}{\partial\mathbf{v}}\cdot\nabla U_j^* + \frac{\partial\rho_j^*}{\partial\mathbf{v}}\cdot\nabla U_i
            \right)
    \right],
    \quad i\ne j.
\end{equation}
\end{widetext}
\par

Thus, from  Eqs. (\ref{f6}), (\ref{f7}), taking into account Eqs. (\ref{a3})--(\ref{a5}), one can see that the induced dipole moment and the functions $q_{ij}(\mathbf{r},\mathbf{v},t)$ (characterizing the coherence between $|bi\rangle$ and $|a\rangle$ states) are determined not only by the magnitude of the fields in a given point but also by their gradients (compare with the well-studied case of the two-level atom \cite{Kazantsev1990}); moreover, the contributions into this effect, connected with the coherent part of the optical field, are proportional to the gradients of the light (Stark) shifts of the energy levels. Note that in a particular case when the field is polarized along one of the axes of the Cartesian coordinate system, for example, along $\mathbf{e}_x$: $U_j=\delta_{xj}U$, $V_j=\delta_{xj}V$, and $\gamma\rightarrow 0$ Eqs.(\ref{f6})--(\ref{f8}) are transformed into the averaged (over high frequency oscillations) generalized  Bloch equations of the kinetic theory of RRFs for the two-level atoms in the bihromatic field \cite{Krasnov1994}.



\providecommand{\noopsort}[1]{}\providecommand{\singleletter}[1]{#1}%

\end{document}